\documentstyle[aps,twocolumn,floats,prl]{revtex}
\input epsf
\newcommand{\mfid}{m_{22}} %\left( {m \over 10^{-22} {\rm eV}} \right)}

\begin{document}

\twocolumn[\hsize\textwidth\columnwidth\hsize\csname
@twocolumnfalse\endcsname

\title{Cold and Fuzzy Dark Matter}

\author{Wayne Hu, Rennan Barkana \& Andrei Gruzinov}
\address{Institute for Advanced Study, Princeton, NJ 08540 \\
Revised \today}
\maketitle
\begin{abstract}
    Cold dark matter (CDM) models predict small-scale structure in excess of 
    observations of the cores and abundance of dwarf galaxies.  These problems 
    might be solved, and the virtues of CDM models retained, even without 
    postulating {\it ad hoc} dark matter particle or field interactions, 
    if the dark matter is composed of ultra-light scalar particles ($m\sim 10^{-22}$eV), 
    initially in a (cold) Bose-Einstein condensate, similar to axion dark matter models.
    The wave properties of the dark matter stabilize gravitational collapse 
    providing halo cores and sharply suppressing small-scale linear power.
\end{abstract}
\vskip 0.5truecm
]

{\noindent \it Introduction.---}
Recently, the small-scale shortcomings of the otherwise widely
successful cold dark matter (CDM) models for structure formation have
received much attention (see \cite{SpeSte99,KamLid99,HogDal00,Pee00}
and references therein).  CDM models predict cuspy dark
matter halo profiles and an abundance of low mass halos not seen
in the rotation curves and local population of 
dwarf galaxies respectively. 
Though the significance of the discrepancies is still disputed
and solutions involving astrophysical processes
in the baryonic gas may still be possible (e.g. \cite{vanetal99}), 
recent attention has mostly focused on solutions involving the dark matter
sector.

In the simplest modification, warm dark matter ($m\sim {\rm keV}$) replaces CDM
and suppresses small-scale structure by free-streaming 
out of potential wells \cite{HogDal00}, 
but this modification may adversely affect structure at
somewhat larger scales. % \cite{WhiCro00}.
Small-scale power could be suppressed more cleanly in 
the initial fluctuations, perhaps 
originating from a kink in the inflaton potential \cite{KamLid99}, 
but its regeneration through non-linear gravitational
collapse would likely still produce halo cusps 
\cite{Mooetal99}.

More radical suggestions include strong self-interactions 
either between dark matter particles \cite{SpeSte99} or in the 
potential of axion-like scalar field dark matter \cite{Pee00}.
While interesting, these solutions require self-interactions
wildly in excess of those expected for weakly interacting
massive particles or axions respectively.  

In this {\it Letter}, we propose a solution involving free
particles only.  The catch is that the particles
must be extraordinarily light ($m \sim 10^{-22}$eV) so that
their wave nature is manifest on astrophysical scales.  
Under this proposal, dark matter halos are stable on small
scales for the same reason that the hydrogen atom is
stable: the uncertainty principle in wave mechanics.
We call this dark matter candidate fuzzy cold dark matter (FCDM).

{\vskip 0.1truecm \noindent \it Equations of Motion.---}
It is well known that if the dark matter is composed of
ultra-light scalar particles $m \ll 1$eV, the occupation numbers
in galactic halos are so high that the dark matter behaves
as a classical field obeying the wave equation
\begin{equation}
\Box \phi = m^{2}\phi\,,
\label{eqn:KleinGordon}
\end{equation}
where we have set $\hbar=c=1$.
On scales much larger than the Compton wavelength $m^{-1}$ but
much smaller than the particle horizon, one can employ a Newtonian
approximation to the gravitational interaction embedded in the
covariant derivatives of the field equation and a non-relativistic
approximation to the dispersion relation. It is then convenient
to define the wavefunction $\psi \equiv A e^{i\alpha}$, out
of the amplitude and phase of the field 
$\phi = A \cos(m t - \alpha)$,
which obeys
\begin{equation}
i(\partial_t + {3 \over 2} {\dot a \over a})\psi
= (-{1\over 2 m}\nabla^2+ m\Psi)\psi  \,,
\label{eqn:cosmoschr}
\end{equation}
where $\Psi$ is the Newtonian gravitational potential. For the
unperturbed background, the right hand side vanishes and the
energy density in the field, $\rho = m^2 |\psi|^2/2$, redshifts
like matter $\rho \propto a^{-3}$.

On time scales short compared with the expansion time, the
evolution equations become
\begin{eqnarray}
i\partial_t \psi = (-{1 \over 2 m}\nabla^2+ m\Psi)\psi  \,, \quad
\nabla^2 \Psi = 4\pi G\delta\rho \,.
\label{eqn:schro}
\end{eqnarray}
Assuming the dark matter also
dominates the energy density, we have $\delta \rho = m^2 \delta|\psi|^{2}/2$.
This is simply the non-linear Schr\"{o}dinger equation for a 
self-gravitating particle in a potential well.  In the particle
description, $\psi$ is proportional to the wavefunction of each particle
in the condensate.  

{\vskip 0.1truecm\noindent \it Jeans / de Broglie Scale.---}
The usual Jeans analysis tells us that when
gravity dominates there exists a growing mode 
$e^{\gamma t}$ where $\gamma^2 = 4\pi G\rho$; however
a free field oscillates as $e^{-i E t}$ or $\gamma^2
= -(k^2/2m)^2$.  In fact, $\gamma^2 = 4\pi G\rho -(k^2/2m)^2$ 
and therefore there is a Jeans
scale 
\begin{eqnarray}
r_J &=& {2 \pi / k_{J}} = \pi^{3/4} (G\rho)^{-1/4} m^{-1/2}\,,
\nonumber\\
    &=& 55 \mfid^{-1/2} (\rho/\rho_b)^{-1/4}(\Omega_m h^2)^{-1/4} {\rm kpc} \,,
\end{eqnarray}
below which perturbations are stable and above which they
behave as ordinary CDM. Here $m_{22}=m/10^{-22}$eV and 
$\rho_b=2.8 \times 10^{11}\Omega_m h^2 M_\odot$ Mpc$^{-3}$ 
is the background density.  The Jeans scale is the geometric 
mean between the dynamical scale and the Compton scale (c.f.~\cite{SanWan99,Sin94,PreRydSpe90})
as originally shown in a more convoluted manner by \cite{KhoMalZel85}.

The existence of the Jeans scale has a natural interpretation:
it is the de Broglie wavelength of the ground state of
a particle in the potential well.  
To see this, note that the velocity 
scales as $v \sim (G \rho)^{1/2} r$ so that the 
de Broglie wavelength $\lambda \sim (m v)^{-1} \sim m^{-1} 
	(G\rho)^{-1/2} r^{-1}$. Setting $r_J=\lambda=r$,
returns the Jeans scale.  Stability below the
Jeans scale is thus guaranteed by the uncertainty principle:
an increase in momentum opposes any attempt to confine the
particle further. 

The physical scale depends weakly on the density, but 
in a dark matter halo $\rho$ will be much larger than the
background density $\rho_b$.  Consider the density
profile of a halo of mass $M $ $[\equiv
(4 \pi r_v^3/ 3) 200 \rho_b$, in terms of the virial
radius $r_v$]  found in CDM simulations
\cite{NavFreWhi96}
\begin{eqnarray}
\rho(r,M) 
&\sim&  {200 \over 3} {f \rho_b \over (c r/r_v) (1+c r/r_v)^2}  \,, 
\label{eqn:profile}
\end{eqnarray}
where
$f(c)  = {c^3 /[\ln(1+c)-c/(1+c)]}$ and 
the concentration parameter $c$ depends weakly on mass.
This profile implies an $r^{-1}$ 
cusp for $r < r_v/c$ which will be
altered by the presence of the Jeans scale. 
Solving for the Jeans scale in the halo $r_{J {\rm h}}$ as a function of
its mass using the enclosed mean density yields
\begin{equation}
r_{J{\rm h}} \sim 3.4 (c_{10}/f_{10})^{1/3} \mfid^{-2/3} M_{10}^{-1/9} 
(\Omega_m h^2)^{-2/9} {\rm kpc}\,,
\end{equation}
where we have scaled the mass dependent factors
to the regime of interest
$c_{10}=c/10$, $f_{10}=f(c)/f(10)$, and $M_{10} = M/10^{10} M_\odot$. 
For estimation purposes,
we have assumed $r_{J{\rm h}} \ll r_v/c$ which is technically violated for 
$M_{10} \lesssim 1$ and $m = 10^{-22}$ eV, 
but with only a mild effect.  
In the smallest halos, the Jeans scale is above the turnover radius 
$r_v/c$, and there is no region where
the density scales as $r^{-1}$.  The maximum circular 
velocity will then be lower than that implied by eqn.~(\ref{eqn:profile}).
More massive halos
will have their cuspy $r^{-1}$ behavior extend from
$r=r_v/c$ down to $r_{J {\rm h}}$.

These simple scalings show that the wave nature of dark matter
can prevent the formation of the kpc scale cusps and substructure
in dark matter halos if $m \sim 10^{-22}$eV. However, 
alone they do not determine what does form instead.  
To answer this question, 
cosmological simulations will be required and this lies beyond
the scope of the present work.  Instead we provide here the tools
necessary to perform such a study, a discussion of possible
astrophysical implications and illustrative one dimensional
simulations comparing FCDM and CDM.

{\vskip 0.1truecm \noindent\it Linear Perturbations.---}
The evolution of fluctuations in the linear regime 
provides the initial conditions for cosmological simulations 
and also directly affects the abundance of dark matter
halos.  Because the initial conditions are set while the
fluctuations are outside the horizon, we must generalize
the Newtonian treatment above to include relativistic effects.

Following the ``generalized dark matter'' (GDM) approach of
\cite{Hu98}, we remap the equations of motion for the
the scalar field in equation~(\ref{eqn:KleinGordon}) onto
the continuity and Euler equations of a relativistic 
imperfect fluid.  First, note that 
in the Newtonian approximation, the current density ${\bf j}
\propto \psi^* \nabla \psi - \psi \nabla \psi^*$ plays the role
of momentum density so that ``probability conservation''
becomes the continuity equation.  
The dynamical aspect of equation (\ref{eqn:cosmoschr}) then 
becomes the Euler equation for
a fluid with an effective sound speed 
$c_{\rm eff}^2 = k^2/4 a^2 m^2$,
where $k$ is the comoving wavenumber.  

This Newtonian
relation breaks down below the Compton scale which for
any mode will occur when $a < k/2 m$.  
In this regime, the scalar field is slowly-rolling in its potential 
rather than oscillating and it behaves like a fluid
with an effective sound speed $c_{\rm eff}^2 = 1$ \cite{Hu98}.  
For our purposes, 
it suffices to simply join these asymptotic solutions and treat the
FCDM as GDM with
\begin{equation}
c_{\rm eff}^{2} = {\cases{ 
        1\,,                   & $a \le k/2m$\,, \cr
	k^{2}/4 a^{2} m^{2}\,, & $a > k/2m$\,, }}
\end{equation}
with no anisotropic stresses in linear theory.
We have verified that the details of this matching have a negligible
effect on the results.
Since the underlying treatment is relativistic, 
this prescription yields a 
consistent, covariant treatment of the dark matter inside and
outside the horizon. The linear theory equations including
radiation and baryons are then solved in the usual way but
with initial curvature perturbations in the radiation 
and no perturbations in the FCDM \cite{ic}. 

The qualitative features of the solutions are easily understood.  
The comoving Jeans wavenumber scales with
the expansion as $k_J \propto a \rho_b^{1/4}(a)$ or
$\propto a^{1/4}$ during matter domination and
constant during radiation domination. 
Because the comoving Jeans scale is nearly constant, perturbation
growth above this scale generates a sharp break
in the spectrum.
More precisely, the critical scale is $k_J$ at
matter-radiation equality
%\begin{equation}
$
k_{J \rm eq} = 9\, \mfid^{1/2}\, {\rm Mpc}^{-1}\,.
$
%\end{equation}
Numerically, we find that the linear density power spectrum
of FCDM is suppressed relative to the CDM case by
\begin{equation}
P_{\rm FCDM}(k) = T_{\rm F}^2(k) P_{\rm CDM}(k) \,,\,\,\, 
T_{\rm F}(k) \approx {\cos x^3 
\over 1+ x^8}\,,
\end{equation}
where $x = 1.61\, \mfid^{1/18} k/k_{J {\rm eq}}$.
The power drops by a factor of 2 at
\begin{equation}
k_{1/2} \approx 
{1 \over 2} k_{J \rm {eq}}m_{\rm 22}^{-1/18} = 4.5 m_{\rm 22}^{4/9}\, {\rm Mpc}^{-1}\,.
\label{eqn:halfpower}
\end{equation}
The break in $k$ is much sharper than those expected
from inflation \cite{KamLid99} or quartic self-interaction 
of a scalar field.  In the latter, the Jeans
scale is fixed in physical 
coordinates so that the suppression is spread over 3-4 orders
of magnitude in scale \cite{Pee00}.  

{\vskip 0.1truecm \noindent\it Low Mass Halos.---} In the CDM model,
the abundance of low mass halos is too high when compared with the luminosity
function of dwarf galaxies in the Local Group
\cite{Klyetal99,Mooetal99b}.  Based on analytic scalings, Kamionkowski
\& Liddle \cite{KamLid99} argued that a sharp cutoff in the initial
power spectrum at $k=4.5 h\, {\rm Mpc}^{-1}$ might solve this problem. Thus,
the FCDM cutoff at $k\sim 4.5\, {\rm Mpc}^{-1}$, produced if $m \sim
10^{-22}$eV is chosen to remove kpc scale cusps, may solve the low
mass halo abundance problem as well.  Whether the required masses
actually coincide in detail can only be addressed by simulations.

Numerical simulations of CDM with a smooth cutoff in the initial power spectrum
qualitatively confirm 
the analytic estimates but suggest that
a somewhat larger scale may be necessary: half power at  
$k=2\, h\, {\rm Mpc}^{-1}$ reduces the $z=3$
abundance of $10^{10}\, h^{-1}\, M_{\odot}$ halos by 
a factor of $\sim 5$ and the abundance of 
$10^{11}\, h^{-1}\, M_{\odot}$ halos by a factor of
$\sim 3$ \cite{WhiCro00}. 
Note, however, that our model produces a much sharper
cutoff in the power spectrum than in the model tested. 
Furthermore, astrophysical influences such as feedback or 
photoionization may have prevented dwarf galaxy halos from 
accumulating much gas or stars \cite{Feedback}.

{\vskip 0.1truecm \noindent\it First Objects and Reionization.---}
At very high redshift, much of the star formation in a CDM model
is predicted to occur in low-mass halos which are not present
in the FCDM model. 
In a CDM model the first round of star formation is thought
to occur in objects of mass $\sim 10^5\, M_\odot$ (see
\cite{HaiThoLoe96} and references therein) due to molecular
hydrogen cooling. 
The consequent destruction of molecular hydrogen \cite{molecules,HaiAbeRee99}
implies that it is larger mass objects
$\gtrsim 10^8\, M_\odot$, where atomic cooling is possible, that are responsible for reionization. 
In our scenario, 
if the cutoff scale in eqn.~(\ref{eqn:halfpower}) were set to reduce the
abundance of $M\lesssim 10^9\, M_\odot$ halos, reionization could be
delayed and the number of detectable galaxies prior to reionization
reduced by a factor of $\sim 5$ \cite{BarLoe00}.

%As for reionization, most of the relevant star formation  occurs
%in more massive galaxies, with virial temperatures $\gtrsim 10^4\,$K, where cooling due to
%atomic transitions is possible. The corresponding minimum mass of
%these objects at $z\sim 10$ is 
%After
%reionization, the substantial increase in the gas
%temperature leads to a suppression of infall and, therefore, star
%formation in the lowest-mass halos (e.g. \cite{Gne00}).
%Thus, if most of the universe remained neutral down to 
%$z\sim 10$, then the standard scenario predicts the existence of a
%substantial population of faint galaxies which are potentially
%observable \cite{BarLoe00}. 

{\vskip 0.1truecm\noindent \it Live Halos.---}
Precisely what effect the Jeans (de Broglie) scale has on the structure
and abundance of low mass halos is best answered through simulations.  
To provide some insight on these issues we conclude
with simulations of the effects in one dimension. 

We solve the wave equation (3) in an interval $0<x<L$ with boundary 
$\psi(0)=\psi(L)=0$. At $t=0$, the density perturbation 
is $\delta \rho = \rho_0 \sin (\pi x/L) \gg \rho_b$, 
with $\psi$ real. We define the Jeans length $r_J$ by equation (4)
with the density $\rho_0$; then the  choice of $r_J/L$ specifies $m$.  
It is also convenient to define the dynamical time scale
$t_{\rm dyn}=(4\pi G \rho_{0})^{-1/2}$.
A one dimensional CDM simulation with 
the same initial density and zero initial velocity was run for comparison.

%However, since the initial field state contains several low-energy levels, 
%interference effects cause evolution on a characteristic time scale 
%$t_{\rm dyn} (L/r_J)^2$.
% $\sim mL^2$  % /\hbar$.  

\begin{figure}[htb]
\centerline{\epsfxsize=3.4truein\epsffile{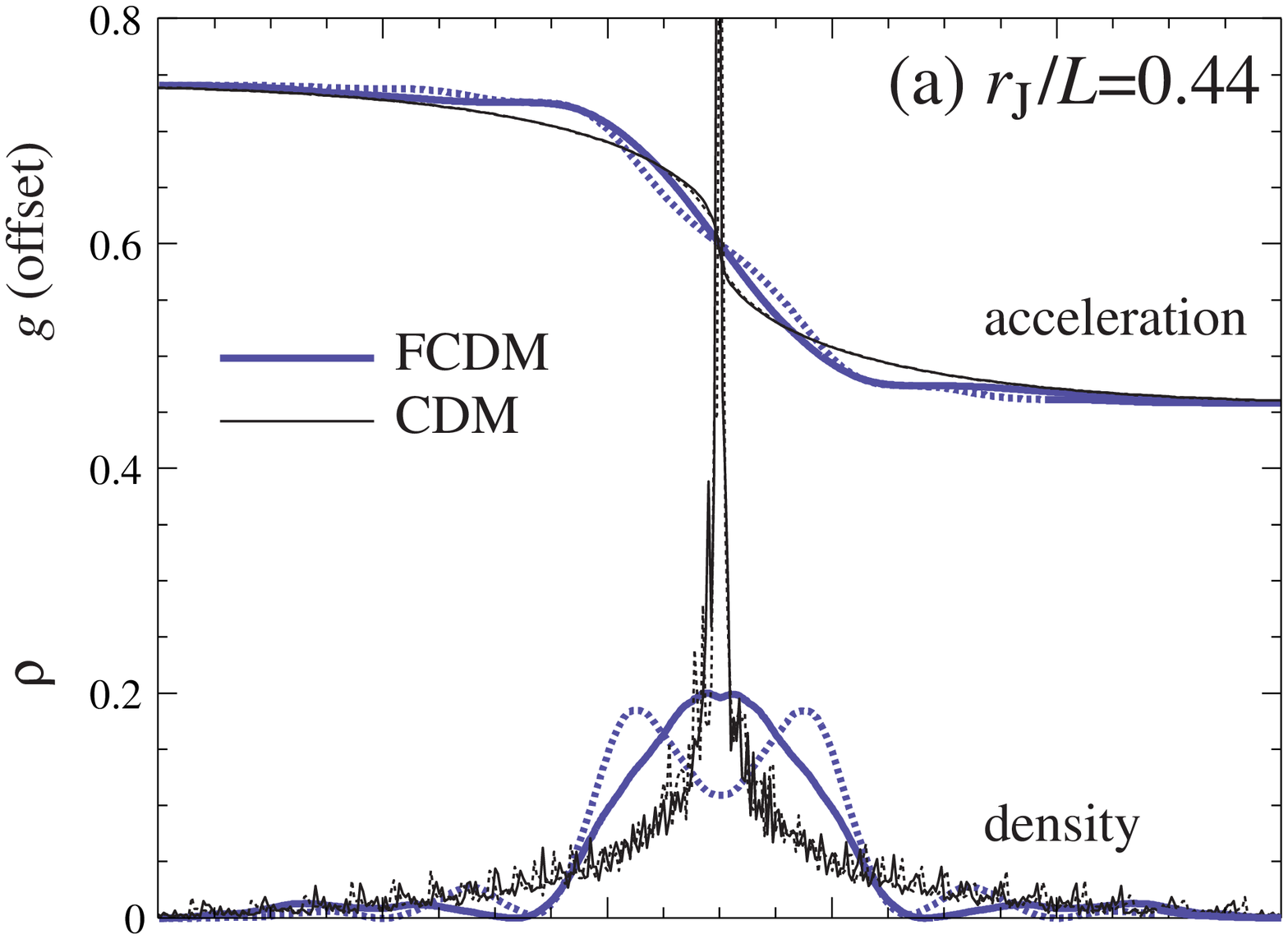}}
\centerline{\epsfxsize=3.4truein\epsffile{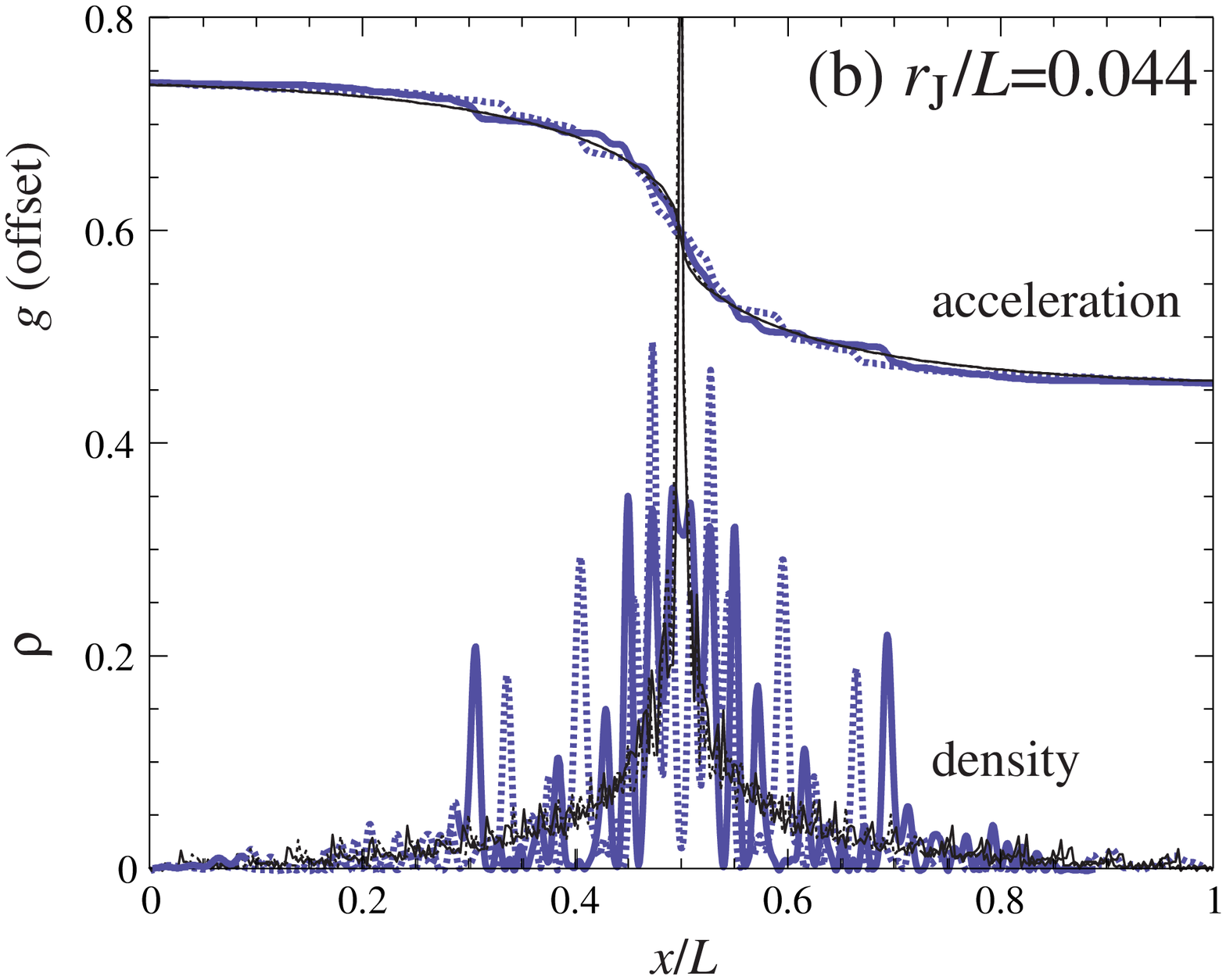}}
\caption{
One dimensional simulations (a) large Jeans scale $r_J/L=0.44$ 
(b) small Jeans scale $r_J/L=0.044$. 
Two snapshots, $t/t_{\rm dyn}=99$ (solid) and 100 (dotted) are shown for the density
profile (units of 15$\rho_0$) and gravitational acceleration
(units of $3 L/ 2 t_{\rm dyn}^2$, offset for clarity).}
\label{fig:sim}
\end{figure}

For $r_J\gg L$, the field model does not form a gravitating halo. 
For $r_J \sim L$ a gravitationally bound halo is formed, but the cusp, 
which is clearly seen in the CDM simulation, is not observed (see 
Fig.~\ref{fig:sim}). Interference effects
cause continuous evolution on the dynamical time scale
$t_{\rm dyn}$ (c.f. \cite{Sin94}).   Note however that the gravitational 
acceleration is much smoother so that the trajectories of test particles
(i.e. visible matter) will be less affected by these wiggles.

For $r_J\ll L$, the density computed from the field follows the CDM
simulation when smoothed over many Jeans scales.
This is to be expected because 
in this limit the Schr\"odinger equation can be solved in the geometrical optics 
approximation. Nonetheless, interference 
features localized in space $(\delta x \sim r_J)$ 
and time $(\delta t \lesssim t_{\rm dyn})$ are quite strong, of order 1. Again
these small-scale features make only a small contribution to 
the gravitational 
acceleration.

{\vskip 0.1truecm\noindent \it Discussion.---}
We have shown that the wave properties of ultra light 
($m \sim 10^{-22}$eV) dark matter can suppress kpc scale
cusps in dark matter halos and reduce the abundance of low mass halos.
While such a mass may seem unnatural
from a particle physics standpoint
\cite{Car98}, even lighter scalar fields
($m \lesssim 10^{-33}$ eV, where the wave nature is manifest across
the whole particle horizon) have been proposed to provide
a smooth energy component to explain observations of
accelerated expansion \cite{quintessence}.

Three dimensional numerical simulations are
required to determine whether our proposal 
works in detail.  Our one dimensional simulations 
suggest that the small-scale cutoff appears at $r \sim r_J$
and that the density
profile on these scales not only is not universal but 
also evolves continuously
on the dynamical time scale (or faster) due to interference
effects.  The observable rotation curves are smoother than the
density profile so that this prediction, while testable with high-resolution
data, is not obviously in conflict with the data today.  Likewise, the
time-variation of the potential is smaller than that of the density but can in 
principle transfer energy from the FCDM to the baryons in the halo.  This
could puff up the baryons in dwarf galaxies while bringing the FCDM closer
to a stationary ground state, but the precise evolution requires detailed 
calculations.

Our Jeans scale is a weak function of density, $r_J \propto \rho^{-1/4}$. 
This has two testable consequences.  
The first is a sharp cutoff at $k \sim 4.5\, m_{22}^{1/2} {\rm Mpc}^{-1}$ in the linear power.
Quantities related to the abundance of low mass halos, e.g., 
dwarf galaxies in the Local Group, the first objects, faint galaxies at very high redshifts, 
and reionization can be seriously affected by the cutoff.
Counterintuitively, quantities related to the {\it non-linear} power spectrum of the dark matter
are only weakly affected due to the
gravitational regeneration of small-scale power \cite{WhiCro00}.
The second consequence is that choosing the mass to set the core radius
in one class of dark matter dominated objects sets the core radius 
in another set, given the ratio of characteristic densities.  
This relation can in principle be tested by comparing the local dwarf 
spheroidals \cite{Mat98} to higher mass systems. 

While the detailed implications remain to be worked out, fuzzy
cold dark matter provides an interesting means of
suppressing the excess small-scale power that plagues the cold dark
matter scenario.

{\vskip 0.1truecm\noindent \it Acknowledgements.---}
We thank J.P. Ostriker, P.J.E. Peebles, D.N. Spergel, M. Srednicki, and M. White for useful discussions.  WH and AG are supported
by the Keck Foundation and NSF-9513835; RB by Institute Funds and the NSF 
under Grant No. PHY94-07194.

\end{document}